\documentclass[twocolumn,showpacs,preprintnumbers,amsmath,amssymb]{revtex4}
\usepackage{graphicx}% Include figure files
\usepackage{dcolumn}% Align table columns on decimal point
\usepackage{bm}% bold math

\begin{document}

\preprint{}

\title{Is flat rotation curve a sign of cosmic expansion?}% 

\author{F. Darabi}
\altaffiliation{Department of Physics, Azarbaijan Shahid Madani University, 53714-161 Tabriz, Iran}
\email{f.darabi@azaruniv.edu}
%\affiliation{%
%uthors' institution and/or address
%his line break forced with \textbackslash\textbackslash}%

%\affiliation{
%econd institution and/or address\\
%his line break forced% with \\
%%

\date{\today}% It is always \today, today,

\begin{abstract}
Four alternative proposals as the possible solutions for the rotation curve problem are introduced on the basis of assumption that the cosmic expansion is engaged with the galactic dynamics over the halo. The first one proposes a modification of equivalence principle in an accelerating universe. The second one proposes a modification of Mach principle in an expanding universe. The third one proposes a dynamics of variable mass system for the halo in an expanding universe, and the fourth one proposes the replacement of physical radius with comoving one over the halo, in an expanding universe.  
\end{abstract}

\maketitle

\section{Introduction}

Observations on the rotation curves of galaxies have turn out that they are not rotating in the same manner as the Solar System, according to the classical Newtonian dynamics. Galactic rotation curves illustrating the velocity of rotation versus the distance from the galactic center, cannot be explained by the luminescence matter. This suggests that either a large amount of the
mass of galaxies is contained in the dark galactic halo, or Newtonian dynamics does not apply universally.

The dark matter proposal is mostly referred to Zwicky \cite{Zwicky}
who found that the motion of the galaxies of the clusters induced by the gravitational field of the cluster can be explained by the assumption of dark matter in addition to the total visible matter content of the observed galaxies. Later, it turned out that dark matter is not the specific property of clusters, rather it can be found as well in single galaxies to explain their flat rotation curves.

The second proposal resulted in the modification of Newtonian dynamics
(MOND) \cite{MOND}. The modification proposed by Milgrom was the
following
\begin{equation}\label{1}
F=m\mu(\frac{a}{a_0})a,
\end{equation}
\begin{equation}\label{2}
\mu(x)=\left \{ \begin{array}{ll} 1 \:\: \mbox{if}\:\: x\gg 1
\\
x \:\: \mbox{if}\:\: \|x|\ll 1,
\end{array}\right.
\end{equation}
where $a_0=1.2\times10^{-10} ms^{-2}$ is a proposed new constant.
At the galactic scale outside the central bulge $a \sim a_0$, so we have the modified dynamics
\begin{equation}\label{0}
F=m(\frac{a^2}{a_0}). 
\end{equation}
Using this new law of dynamics for the gravitational force we obtain
\begin{equation}\label{3}
\frac{GM}{r^2}=(\frac{a^2}{a_0}),
\end{equation}
where $M$ is the total mass of the bulge. 
This results in  
\begin{equation}\label{4}
v^2=\sqrt{GMa_0}.
\end{equation}
Alternative solutions in this direction has also been proposed by assuming the gravitational attraction force to be $1/r$ beyond some galactic scale \cite{Sanders}, or assuming the dark matter as the manifestation of Mach principle showing the effective gravitational mass of astrophysical bodies to be $r$ dependent \cite{Treder}. Recently, some authors have addressed
the rotation curve problem by resorting to modified theories of gravity \cite{Lin},
\cite{Capo}. 

In what follows we will introduce four alternative proposals for the rotation curve problem without resorting to the dark matter, MOND or modified gravity. These proposals are based on the possibility that the cosmic expansion or
acceleration is next to the galactic structures and the dynamics of the outer (halo) structure of the galaxies is engaged with the cosmological dynamics.
The results of the first proposal for the rotation curve is in good agreement with that of obtained recently by Lin {\it et al.} \cite{Lin} based on the Grumiller's modified gravity \cite{Grum}.

\section{Modified equivalence principle due to cosmic acceleration}

Recent cosmological observations obtained by SNe Ia {\cite{c1}}, WMAP {\cite{c2}}, SDSS {\cite{c3}} and X-ray {\cite{c4}} indicate that our universe is globally
experiencing an accelerated expansion. This acceleration is imposed on empty space all over the universe, so that for an observer in a local rest frame, the space within the frame is isotropically expanded radially outward in
an accelerating way, namely we have an isotropic acceleration of space around each point in this rest frame. 

Let us now revisit the {\it equivalence principle} in the presence of this {\it space acceleration}. We know that in a static space, the equivalence principle states that a local frame rested in the vicinity of a gravitational field with gravitational strength $g$ is indistinguishable from an accelerated frame (in empty static space) having an acceleration $a_{out}$ equal to the gravitational strength $g$. The equivalence principle is also the main reason by which, for example, a massive body rotates around a gravitational mass $M$ over a constant radius $r$ as
\begin{equation}\label{4'}
\frac{v^2}{r}=\frac{GM}{r^2},
\end{equation}
where the centrifugal and the gravitational accelerations play the role of $a_{out}$ and $g$, respectively.

However, the story is different in a frame in which the background space is not static. Consider a free local frame 1 in such a expanding space. If we put a body of mass $m$ in this frame, then it experiences a radially outward and isotropic cosmic acceleration in all directions, so the overall acceleration imposed on the body vanishes. Now, let us impose an upward directional acceleration on this frame by an outer agent. It destroys the isotropy of space from the observer's point of view and a local acceleration is induced in this frame. This is similar to the spontaneous symmetry breaking phenomena where the system after symmetry breaking takes a ground state. For example, the interactions between the atoms in a ferromagnet is invariant under rotation. However, when the temperature reaches the critical temperature the rotational symmetry disappears and a ground state appears in which all the spins are aligned, and this is clearly not rotationally invariant. In our case, before the outer
agent of acceleration is applied on the frame, the space within the frame has isotropic symmetry so that a massive body therein experiences a same acceleration in all radial directions and so the resultant force imposed
on this body vanishes. The symmetry disappears when a preferred direction is singled out by an outer agent which imposes an acceleration in a preferred upward direction on the frame. Therefore, a local acceleration (as a singled
out ground state) emerges within the frame which is clearly not isotropically invariant. It is reasonable to suppose that the direction of the induced acceleration within the frame is in the same direction of the upward acceleration, imposed by the outer agent, which breaks down the isotropy. Hence, if the outer upward acceleration of the frame
is $a_{out}=g$, the body inside this local frame bears an upward induced acceleration $a_{c}$, despite the downward acceleration $g$. Note that, the induced acceleration $a_{c}$ should depend just on the value of the cosmic acceleration, hence we call it {\it induced cosmic acceleration} $a_{c}$.

Now, in order to validate the equivalence principle, we assume a local frame 2 in the vicinity of a gravitational field, with strength $g$, in a static background space and put a body of same mass $m$ in this local frame. In this frame, the body falls down towards the gravitational center by an acceleration $g$. But, in the previous paragraph we realized that the body of mass $m$ in the frame 1 (under the outer acceleration $a_{out}=g$) does not fall with the acceleration $g$, rather it falls with an acceleration as the resultant combination of $g$ and $a_{c}$. It is clear that the discrepancy between the observations in the frames 1 and 2 is due to the fact that the background space is not static within the frame 1. Once we take a static space within the frame 1, both masses in the frames 1 and 2 fall with the same acceleration
$g$, so the discrepancy is removed and the equivalence principle is
reestablished. However, it is still possible to reestablish the equivalence principle between two frames 1 and 2, even if the space within the frame 1 is not static. To this end, the outer acceleration imposed on the frame 1 should first compensate the induced cosmic acceleration $a_{c}$ (acceleration of the background space itself) within this frame and then provide the extra acceleration equal to $g$ (by which the body falls down in the frame 2) as 
\begin{equation}\label{5}
a_{out}=a_{c}+g,
\end{equation}
which gives a resultant acceleration in the frame 1, as 
\begin{equation}\label{5'}
\bar{a}=a_{out}-a_{c}=g.
\end{equation}
Note that, the value of $a_{c}$ may depend on the cosmic acceleration
and the value of $a_{out}$ imposed on the frame 1. This is what we mean by {\it modified equivalence principle}. According to this equation, both masses in the frames 1 and 2 now fall downwards with the same acceleration $g$.

Now, suppose a galaxy with a dens mass in the bulge and a diluted
mass in the halo. We assume that, at present state of low
accelerating phase, the cosmic acceleration is just enable to affect the halo and not the tight bound bulge. However, when the universe enters the phantom phase, not only the bulge but also all the gravitational structures, the electromagnetic structures, and even the strongly coupled structures will be thorn apart by the cosmic acceleration. 

Let us now consider a typical galaxy with a bulge of radius $R_0$ and a body of mass $m$ located at a radius $r\gtrsim R_0$ in the halo within a local frame rotating around the center of galaxy. This body experiences a real gravitational force $F={GmM}/{r^2}$ where $M$ is the total mass of the bulge and $r$ is the distance from the center of the galaxy. However, it experiences a centrifugal force as well, due to the rotation. Now, let us assume that the cosmic acceleration is next to the galaxy and so affects the halo. Using the above modified equivalence principle, we may consider the rotating frame here as: 1) the frame 1 concerning the centrifugal acceleration as $a_{out}$, and 2) the frame 2 concerning the gravitational acceleration as $g$. According to the above discussion, the centrifugal force as the agent of outer acceleration breaks the space isotropy within the frame 1 and results in a upward induced cosmic acceleration $a_i$ inside this local frame. Therefore, using the equivalence of inertial and gravitational masses $m_i=m_g$, we may rewrite the modified
equivalence principle Eq.(\ref{5}) as 
\begin{equation}\label{6}
m_i\frac{v^2}{r}=m_ia_{c}+\frac{Gm_gM}{r^2}.
\end{equation}
Note that the induced cosmic force $m_ia_{c}$ which is upward from the observer's point of view in the rotating frame considered here, is actually trying to pull out radially the body of inertial mass $m_i$ from the halo's {\it gravitational configuration} towards the {\it cosmological void background}. One may interpret the induced force $m_ia_{c}$
in the right hand side of Eq.(\ref{6}) as the reaction force which the gravitational system of the galaxy imposes against the cosmic force attempting to pull out the body of mass $m$ from halo. Eq.(\ref{6}) results in the following expression for the orbital velocity within the halo 
\begin{equation}\label{7}
{v}=\sqrt{ra_{c}+\frac{GM}{r}}.
\end{equation}
Note that since the bulge is not affected by the cosmic
acceleration, the velocity profile within the bulge is Keplerian 
\begin{equation}\label{7''}
{v}=\sqrt{\frac{GM(r)}{r}}.
\end{equation}
However, if we assume the bulge is also affected by the cosmic
acceleration, then we have
\begin{equation}\label{}
{v}=\sqrt{ra_{c}+\frac{GM(r)}{r}},
\end{equation}
for $r>0$.

Now, we may ask the interesting questions:\\ {\it - What is the interpretation of the ``outer agent"?}\\ {\it - Why the induced cosmic acceleration is simply upward?}\\ To answer the first question, we note that in principle the outer agent is just meaningful in the framework of the equivalence principle in general relativity. In a thought experiment, when the ``outer agent" accelerates a rest frame by $a_{out}$, a same amount of pure gravitational acceleration $g=a_{out}$ is effectively produced inside the frame, exerting on a test body. In the framework of our generalized equivalence principle, the ``outer agent" accelerates a rest frame by $a_{out}$ and a pure gravitational acceleration $g$ plus an induced cosmic acceleration $a_{c}$, originating from the existence of cosmic acceleration in space and breakdown of space isotropy, is effectively produced inside the frame, exerting on the test body. 

In fact, the outer agent does not really exist in these thought experiments, but helps us to realize the equivalence between the inertial and gravitational forces: {\it Wherever there is a gravitational force, we may look for the corresponding equivalent inertial force}. In the present paper, we are looking for the origin of a gravitational force which produces a flat rotation curve within the halo. Hence, we resort to the (generalized) equivalence principle to find the origin of the extra gravitational force which causes deviation from the Keplerian curve. We find that the origin of this extra gravitational force on a test body within the halo may be nested in the equal inertial force exerting by an induced cosmic acceleration $a_c$ on this test body in a local frame (within the halo) which is being accelerated by
the thought {\it outer agent} with $a_{out}$. In other words, the {\it outer agent} here is nothing but a thought agent accelerating (radially outward)
a rest frame within the halo to produce a locally equivalent gravitational force (radially inward) inside the frame, exerting on the test body to keep it on a local orbit over a flat rotation curve within the halo.

To answer the second question, we note that {\it upward acceleration} is just a relative concept. A local rest frame which is rotating on an orbit around the center of galaxy, is also a locally free falling frame along a radial direction towards the center of galaxy. Hence, in this frame ``upward" and ``downward"  means ``radially outward" and ``radially inward", respectively. By symmetric considerations, it is reasonable to suppose that the direction of the induced cosmic acceleration exerting on the test body in this local frame is in the same radial direction of the acceleration imposed by the thought outer agent which produces this induced cosmic acceleration. The reason for an upward, instead of downward, acceleration is simply because according to our assumption the space within the halo is radially accelerating outward the bulge, so for a test body (or a rest observer) in the local frame within the halo, the radially outward acceleration means exactly an upward acceleration.

It is appealing to connect the cosmic acceleration here with the acceleration
constant introduced by Milgrom in his theory of MOND. According to Milgrom
\cite{{Mil1}}, the acceleration constant, which marks the boundary between the validity regions of MOND and Newtonian dynamics, turns out to have a value that matches accelerations appearing in the context of cosmology, namely the recently discovered acceleration of the universe which is of the same order of magnitude as $cH_0$ and $c\sqrt{\Lambda/3}$ where $c$, $H_0$ and $\Lambda$ are the velocity of light, Hubble constant, and the emerging value of the cosmological constant, respectively. 
Similar to Milgrom, we take this statement as a hint that Milgrom's acceleration is somehow connected with the current accelerating phase of the universe.  

Motivated by the above idea, we assume that the cosmic acceleration $a_{c}$ is calculable from a deeper theory so that its estimate value
at the edge of bulge coincides with the Milgrom's acceleration
constant $a_0\simeq 1.2\times 10^{-10}ms^{-2}$. This is similar to the {\it constant} $g$, namely the free-fall acceleration near Earth's surface, as it appears, say, in Galilean mechanics. The deeper theory in this case is Newtonian universal gravity, which tells us that $g$ is calculable from the mass and radius of the Earth \cite{Mil1}. Now, putting the value of $a_0$ in Eq.(\ref{7}) gives the following formula for the orbital velocity at  $r\gtrsim R_0$
\begin{equation}\label{8}
{v}\simeq\sqrt{1.2\times 10^{-10}r+\frac{GM}{r}}.
\end{equation}
Applying this formula to a typical galaxy indicates that:\\
 i) the maximum value of velocity at $R_0$ is a little greater than that of Keplerian rotation curve due to the first term including $a_0$,\\
ii) the velocity decreases to a local minimum at a distance $R_0\simeq{GM}/{a_0} $,\\
 iii) the velocity very steadily climbs for $r\gtrsim R_0$.  

Eq.(\ref{8}) results in the rotation curve profile of the halo in good agreement with that of dwarf galaxies, or large (spiral) galaxies, see e.g. \cite{Rhee} and \cite{Sofu}, respectively.
However, since there is no empirical hint that the rotation curves of galaxies rise linearly at large distances, see e.g. \cite{Blok}, so we may assume the induced cosmic acceleration as a function of distance $r$ over the halo,
and rewrite (\ref{7}) as
\begin{equation}\label{7'}
{v}=\sqrt{ra_{c}(r)+\frac{GM}{r}}.
\end{equation}
This is similar to the fact that the gravitational acceleration $g$ at
the surface of Earth with radius $r_0$ is almost constant while in principle it is $r$-dependent for larger radiuses $r>r_0$. Hence, we may assume that the induced cosmic acceleration at the edge of bulge with radius $R_0$ is almost constant but it is $r$-dependent for $r\geq R_0$ in the halo. Actually,
in Eq.(\ref{5'}) we have already assumed the dependence of $a_c$ on $a_{out}$ in the modified equivalence principle. Since $a_{out}=v^2/r$, then we may
expect $a_c=a_c(r)$. Actually, allowing for such a $r$-dependent function one can fit any given rotation curve. In fact, such an $r$-dependence may be well-justified for the induced cosmic acceleration considered here, as follows. 

The induced cosmic acceleration in the void and distant spaces between the galaxies is almost vanishing because in these regions the space is accelerating radially in all directions in an isotropic way, so the resultant acceleration imposed on a test body becomes zero at each arbitrary point in the void and distant spaces between the galaxies. For this reason, very far away from the whole gravitational structure of a typical galaxy, the induced cosmic acceleration is also zero for a test body at any point. This property limits the scope of the induced cosmic acceleration to the galactic halo as follows. Inside the bulge, because of the strong and tight gravitational bound between the gravitational ingredients, there is no engagement of the cosmic acceleration and galactic dynamics. So, there is no room for the induced cosmic acceleration inside the bulge. As is discussed above, well beyond the halo, the induced cosmic acceleration is also vanishing due to the full isotropy of space. Therefore, it is just left for the halo to exhibit the engagement of the cosmic acceleration and the galactic dynamics. This means that the induced cosmic acceleration is
non-vanishing and has local property just within the halo. 

For a given mass of a test body at the edge of bulge, namely at the beginning
of the halo, there is a considerable centripetal acceleration radially directed towards the center of galaxy, and this singled out radial direction easily disturbs the isotropy of space at this region. Hence, the resultant induced cosmic acceleration in this region has a rather considerable non-vanishing value along the radial direction. It is therefore plausible that for a given mass at more distant regions along the radial direction within the halo, where the centripetal acceleration becomes smaller, the disturbance on the
isotropy of space should be smaller as well and so the resultant induced cosmic acceleration in this region becomes smaller. In other words, the more we get closer to the gravitational center within the halo, the more centripetal acceleration we have, the more isotropy of space is disturbed, and the more resultant induced cosmic acceleration we have. On the other hand, the more we go away from the gravitational center within the halo, the less centripetal acceleration we have, the less isotropy of space is disturbed, and the less resultant induced cosmic acceleration we have. Hence, the induced cosmic acceleration becomes small and smaller at large and larger distances within the halo respectively, and vanishes in the far void space beyond the galaxy, as we first expected. 

Therefore, the formula (\ref{7'}) with a local $r$-dependent and halo-dependent
induced cosmic acceleration is well justified, and can predict the flat rotation curves by considering the characteristic features of each halo in each galaxy. This may justify the variety of flat rotation curves corresponding to different types of galaxies. A very simple function for the induced cosmic acceleration may be proposed as  
\begin{equation}\label{9}
a_{c}=\frac{R_0}{r}a_0,
\end{equation}  
where $R_0$ is the characteristic size of the bulge. Putting this into Eq.(\ref{7'})
gives
\begin{equation}\label{10}
{v}=\sqrt{R_0a_0+\frac{GM}{r}},
\end{equation}
which produces an almost flat rotation curve for $r>R_0$. Since the cosmic acceleration can not affect the inner bulge $r<R_0$ (with tight gravitational structure), there is no induced cosmic acceleration in the bulge ($a_c=0$) and so the orbital velocity within $r< R_0$ and at $r\simeq R_0$ is obtained respectively
as
\begin{equation}\label{10'}
{v}=\sqrt{\frac{GM}{r}},
\end{equation}
and
\begin{equation}\label{11}
{v_0}\simeq\sqrt{\frac{GM}{R_0}}.
\end{equation}
Using Eq.(\ref{11}), we may replace for $R_0$ in Eq.(\ref{10}) and obtain \begin{equation}\label{12}
{v}\simeq\sqrt{\frac{GM}{v_0^2}a_0},
\end{equation}
where ${GM}/{r}$ is almost ignored for $r>R_0$. This formula coincides
numerically with that of Milgrom, namely (\ref{4}), because numerically we have $v\simeq v_0$. Therefore, another result of the proposal discussed in
this section is that the Milgrom's constant acceleration is nothing but the induced $r$-dependent cosmic acceleration at $r=R_0$, just like $g$ which is nothing but the $r$-dependent gravitational acceleration at the surface of Earth.

The formulas (\ref{8}) and (\ref{7'}) are in close agreement with those suggested
initially by Grumiller \cite{Grum} and discussed afterwards by Lin {\it
et al.} \cite{Lin} in fitting to the rotation curve data of halo for some galaxies. The major difference lies in the physical origins of the extra accelerations appeared in the formulas here and those suggested by Grumiller.
The effective potential in the Grumiller's modified gravity includes the Newtonian potential and a Rindler term, so the extra acceleration suggested by Grumiller has its origin in Rindler term. However, the induced acceleration
in the present work has its origin in the cosmic acceleration

\section{Modified Mach principle and cosmic inertial mass}

According to Mach principle \cite{Mach} the distant mass
distribution of the static universe has been considered as being
responsible for generating the local inertial properties of the
close material bodies. One may modify Mach principle in an expanding universe
so that the local inertial properties of the material bodies are affected by the expansion of universe. In this line of thought, we assume that the inertial mass of a body is constant within the bulge of a typical galaxy where a dense distribution of matter presents. This is in accordance with the spirit of Mach principle in that as long as a large and almost constant mass configuration exists around a body, it has a constant inertial mass. In fact, according to Mach, the coordinates in space are defined by the presence
of matter configurations. In other words, the space does not exist without the presence of matter. Therefore, a deep connection exists between
the inertial properties of matter and the coordinates in space. In this regard,
one may think that within the dense part of the galaxy, namely the bulge,
where all the mass configuration is rotating as a rigid body, the inertial
properties of each individual body, namely the inertial mass, is constant.
This is because the coordinates of bodies (configurations) in this almost
rigid body is constant. So, their inertial properties (mass) which are induced by this constant spatial configuration becomes constant too. In fact, the matter configuration in the bulge is constructed by the gravitational interaction which tightly bounds together all the bodies inside. These tightly bound
(dense) bodies define the tightly bound (dense) spatial coordinates within the space of bulge, and these dense coordinates determines the tight inertial property or {\it constant inertial mass} of bodies in the bulge. 

However, the situation changes for bodies beyond the bulge in the halo where, according to our assumption, they are affected by the expansion of universe. In fact, we assume that the inertial mass of bodies beyond the bulge is diluted because their physical configuration is subject to the expansion of universe. In other words, the spatial coordinate $r$ beyond the bulge, $r>R_0$, is subject to the expansion of universe so that unlike the dense spatial {\it
comoving} coordinates within the bulge, $r<R_0$, the {\it comoving} coordinate $r$ beyond the bulge is not dense, rather it is diluted due to the expansion of universe. Therefore, according to Mach, the inertial property or mass of bodies beyond the bulge may be diluted too, in comparison to the tight inertial property or constant mass of bodies within the bulge. Then, one may suggest the following definitions for the inertial mass within and beyond the bulge, respectively as
\begin{equation}\label{13}
\left\{ \begin{array}{ll} m_{i}=C, ~~~~ r\leq R_0,\\
\\
m_{i}=\frac{C'}{r},~~~~ r>R_0,
\\
\end{array}
\right.
\end{equation}
where $C$ and $C'$ are constants. We call the first one as inertial mass versus {\it gravitational interaction}
within the bulge, and the second one as inertial mass versus {\it cosmological expansion} beyond the bulge. Eq.(\ref{13}) is what we mean by
{\it modified Mach principle}.
 
Now, using (\ref{13}), we may write the equation of motion for a body at
$r>R_0$ as follows 
\begin{equation}\label{14}
m_i\frac{v^2}{r}=\frac{Gm_gM}{r^2} \Rightarrow v=\sqrt{\frac{GM}{R_0}},
\end{equation} 
where $C'=m_gR_0$ is taken based on the dimensional considerations to obtain
the flat rotation curve. This formula again gives a rotation curve in agreement
with observations. Note that it seems at first glance that the equivalence principle is violated in the case of second definition in (\ref{13}). However, this is not the case. In fact, all bodies with different inertial masses at the same radius $r>R_0$ are still falling with a same centripetal acceleration ${v^2}/{r}={GM}/{rR_0}$ toward the center of galaxy while rotating around it.

The general assumption of an $r$-dependent cosmological inertial mass like in (\ref{13}) becomes plausible if we remind the cosmological radial velocity $\dot{r}=Hr$, namely the Hubble law. In fact, such a $r$-dependent cosmological inertial mass is justified if we demand for the conservation of cosmological radial momentum for a typical body of inertial mass $m_i$, in the cosmological background $r>R_0$ and at each given cosmological era with constant Hubble parameter, as follows
\begin{equation}\label{15}
m_i\dot{r}=Const.
\end{equation}

\section{Modified Newton's equation due to cosmic expansion}

It is well known that our universe is expanding and according to Hubble law
a body at distance $r$ from an observer experiences a radial velocity
\begin{equation}\label{16}
v_r=Hr,
\end{equation}
where $H$ is the Hubble parameter. Usually, this expansion is assumed to
be considerable at large scales very much larger than the scale of galaxies.
However, let us suppose this expansion is very close to the galaxies so that
the halos of galaxies are subject to the cosmic expansion. Now, we investigate the effect of this expansion on the galactic halo dynamics. 

To this end, we first remind a simple problem in elementary mechanics, namely a system with variable mass. In mechanics, a variable-mass system is a system which has mass that does not remain constant with respect to time. In such a system, Newton's second law of motion cannot directly be applied, instead, this system can be described by modifying Newton's second law and adding a term to account for the momentum carried by mass entering or leaving the system as
\begin{equation}\label{17}
v_{rel}\frac{dM}{dt},
\end{equation}
where $v_{rel}$ is the relative velocity of the entering or leaving mass $dM$ with respect to the center of mass of the body.

For a moment, suppose we turn off the rotation of galaxy and consider a body of constant mass $m$ beyond the bulge in the halo and under the influence of the gravitational force exerted by the bulge having a mass $M$ as ${GmM}/{r^2}$. Also, we assume the radial cosmic velocity (\ref{16}) to be responsible for escaping the matter from the halo of galaxies towards the outer void space. So, the galaxy's halo becomes a typical example of a variable-mass system with
\begin{equation}\label{18}
v_{rel}=v_r=Hr.
\end{equation}
It is reasonable to suppose that in order for the $v_{rel}$ (namely $Hr$) be constant with respect to time (Hubble parameter is almost constant), the physical power $W$ exerted by the cosmic
expansion on the variable mass system in the halo of the galaxy should be
constant as well. Therefore, we have
\begin{equation}\label{19}
W=(Hr)^2\frac{dM}{dt}=C,
\end{equation} 
where $C$ is a constant. This results in 
\begin{equation}\label{20}
\frac{dM}{dt}=C(Hr)^{-2}.
\end{equation} 
Obviously, we are concerned about the dynamics of the individual mass $m$. But, in principle this mass in the halo of galaxy belongs to the same leaving mass $dM$. Hence, it is plausible to suppose 
\begin{equation}\label{21}
dM \sim m,
\end{equation}
and according to (\ref{19}) we find
\begin{equation}\label{22'}
C \sim m.
\end{equation}
This equation reasonably states that the power exerting by the cosmic expansion on the mass $m$ to pull it out from the halo of galaxy is proportional to the mass $m$ itself. Now, by putting (\ref{18}) and (\ref{20}) into Eq.(\ref{17}) and using (\ref{22'}) we obtain the force exerting by the cosmic expansion on the mass $m$ to pull it out from the halo, as follows
\begin{equation}\label{22}
v_{rel}\frac{dM}{dt}=mC'(Hr)^{-1},
\end{equation}
where $C'$ is another constant. Actually, the mass $m$ mediates this force to the bulge of the galaxy (as the main gravitational source) to which it is gravitationally bound. In fact, the cosmic expansion tends to breakdown the whole gravitational structure of the galaxy by pulling radially out {\it all} the individual masses in the galaxy. Hence, according to the third law of action and reaction, an equal and radially inward force is exerted by the gravitational source of the galaxy, namely the bulge, on the mass $m$ to keep it in the halo structure. 

Now, let us turn on the rotation of galaxy. The modified Newton's equation of motion for the mass $m$ inside the halo while rotating with the orbital velocity $v$ at a distance $r$ from the center of the galaxy is written  
\begin{equation}\label{23}
m\frac{v^2}{r}=\frac{GmM}{r^2}+\frac{mC'}{Hr},
\end{equation}
where the second term in RHS emerges due to the cosmic expansion. Eq.(\ref{23}) leads to the following result
\begin{equation}\label{24}
{v}=\sqrt{\frac{GM}{r}+\frac{C'}{H}}.
\end{equation}
This formula again produces an almost flat rotation curve in the halo $r>R_0$
provided the second constant term in the square root is considerably large.
Being the Hubble parameter in the denominator indicates that at early times
in the universe's age when the young galaxies were constructed, the second
term in the square root was so small that the orbital velocity could be assumed
obeying the Keplerian rotation curve
\begin{equation}\label{25}
{v}\simeq\sqrt{\frac{GM}{r}}.
\end{equation}
However, at the present when the universe is old enough and the Hubble parameter
is small enough, the second term in the square root is considerably large
and can cause for a flat rotation curve.
 
\section{Cosmological scaling}

We know that according to Newtonian dynamics, the rotation curve within the halo is theoretically expected to be Keplerian
\begin{equation}\label{25'}
{v}=\sqrt{\frac{GM}{r}},
\end{equation}
where $M$ is the total mass of the bulge and $r$ is the physical coordinate. However, the observations indicates a flat rotation curve. To solve this problem, we assume that the halo (without dark matter) still obeys the Keplerian dynamics (\ref{25'}). But, on the other hand, we assume that the halo is affected by the cosmic expansion such that the physical coordinate of a body in the halo is $a\bar{r}$ instead of $r$, $a$ being the cosmological scale factor and $\bar{r}$ is the comoving coordinate. In other words, we assume that in the halo the physical coordinate $r$ is replaced by a comoving coordinate $\bar{r}$. 
Therefore, while the rotation curve within the bulge with respect to the
physical radius $r$ is Keplerian (\ref{25'}), the rotation curve in the halo with respect to the comoving coordinate $r$ is Keplerian too 
\begin{equation}\label{27}
{v}=\sqrt{\frac{GM}{\bar{r}}}.
\end{equation}
But, the comoving coordinate ${\bar{r}}$ over the halo is limited to a very small domain by the following condition 
\begin{equation}\label{28}
{R_0}\leq{\bar{r}}\leq {R_0}+\epsilon,
\end{equation}
where $\epsilon$ is very small. Note that $r=\bar{r}=R_0$ corresponds to the boundary of the halo (beginning of the effective cosmological domain) where the comoving coordinate $\bar{r}$ is scaled by the scale factor $a_0=1$, as a boundary value at $R_0$. But, the end of halo denoted by the radius $R_H$ (from the center of galaxy) is scaled as
\begin{equation}\label{29}
R_H\sim ({R_0}+\epsilon)a_H,
\end{equation}
where $a_H$ denotes the value of scale factor at $\bar{r}={R_0}+\epsilon$. In explicit words, the {\it apparent} radial coordinate $r$ in the halo, as the physical distance from the center of galaxy, may be nothing but a comoving coordinate ${\bar{r}}$ limited to a very small domain (see (\ref{28})) which is scaled by a ${\bar{r}}$-dependent expansion factor $a(\bar{r})$ over the halo. In fact, we assume that the scaling rate of the space in the halo (with massive structure) as $a(\bar{r})$ is different from the fixed
scaling rate of the empty space (with no structure) in the universe, namely
$a$. Using (\ref{28}), (\ref{29}), and $r={\bar{r}}a(\bar{r})$ we obtain
\begin{equation}\label{30}
{R_0}\leq{r}\leq {R_H}.
\end{equation}
Taking the variation of both sides of (\ref{27}) leads to
\begin{equation}\label{31}
|\Delta v|=\frac{v}{2\bar{r}}\Delta \bar{r}.
\end{equation}
Inserting $\Delta \bar{r}=\epsilon$ and $\bar{r}\simeq R_0$ from (\ref{28}) leads to    
\begin{equation}\label{33}
|\Delta v|\simeq 0,
\end{equation}
which accounts for a flat rotation curve throughout the halo.

\section{Conclusion}

In this paper, we have given four proposals each of which is
theoretically a possible solution for the rotation curve problem.
Our motivation was the assumption that the rotation
curve problem may be related to the cosmic expansion which
is next to the galactic structure, rather than the dark matter,
MOND or modified gravity. In explicit words, we have
tried to answer the important questions: Can cosmic dynamics
influence anyway the galactic dynamics? Do we expect at
all that the galactic dynamics be potentially engaged with
the cosmic dynamics? Is it time to consider such a engagement
after billions of years, or we should still wait for the
universe to cross the phantom divide and experience a considerable
acceleration? If the answers to the above questions
is positive, then we should concern about this subject seriously.
In our opinion, such a engagement is inevitable and
soon or late it happens. In this study, we have tried to show
the possibility that the galactic flat rotation curve may be
a sign and direct consequence of such a engagement.

In the first proposal, by a generalization of equivalence
principle in an accelerating universe, we showed that the
current acceleration of the universe may be responsible for
the flat rotation curve. The important point is that, even if
this proposal is not correct for the justification of flat rotation
curve, its consequences must be considered seriously
when the universe enters the phantum phase of acceleration
and the halo is inevitably affected by the considerable acceleration
of this phase. In other words, if the flat rotation
curve has its real origin in the dark matter, MOND, or modified
gravity, one should concern about the impact of cosmic
acceleration on the dynamics of halo (and even bulge) when
the universe undergoes the phantum phase acceleration. The
first proposal, at least predicts a modified dynamics of halo
due to the possible engagement with the cosmic dynamics,
at present or in the future at phantom era. The velocity
profile obtained in this proposal is very much like that of
suggested initially by Grumiller \citep{Grum} and discussed later by Lin {\it et al.} \citep{Lin} in fitting to the rotation curve data of halo for some galaxies, except for the different
origins of the extra acceleration which is appeared in
Grumiller model as Rindler acceleration, and here as cosmic
acceleration. It is appealing to investigate the possible relation
between Rindler acceleration and cosmic acceleration
from the present point of view.

In the second proposal, we discussed on the possible
impact of Mach principle on the dynamics of halo in an
expanding universe. We showed that the dilution of inertial
mass, as a consequence of Mach principle in an expanding
universe, may produce a flat rotation curve over the halo.

In the third proposal, we assumed the halo as a variable
mass system whose mass is subject to the cosmic expansion
which is trying to pull the massive bodies out of the halo.We
showed that the reaction force of the gravitational structure
in the galaxy against the pulling out force of cosmic expansion
can cause for a flat rotation curve.

Finally, in the fourth proposal, we assumed that the
halo obeys the Keplerian dynamics with respect to the comoving
coordinate, rather than the physical radius. Since
the comoving coordinate in the halo is limited to a very
small domain, the Keplerian dynamics predicts an almost
constant rotation velocity with a very small variation over
the halo.

\section*{Acknowledgment}
The author would like to thank the anonymous referee for the enlightening comments.

\end{document}